\documentclass[12pt]{article}

\usepackage{scicite}

\usepackage{times}

\usepackage{amsmath}
\usepackage{amsfonts}
\usepackage{amssymb}
\usepackage{graphicx}
\usepackage{xcolor}
\usepackage{hyperref}
\usepackage{clipboard}
\usepackage{comment}
\usepackage[detect-all,group-separator={,}]{siunitx}

\newif\iffinal

\iffinal
  \newcommand\logan[1]{}
  \newcommand\andre[1]{}
  \newcommand\ian[1]{}
  \newcommand\ben[1]{}
  \newcommand\CWL[1]{}
\else
  \newcommand\logan[1]{{\color{violet}[Logan: #1]}}
  \newcommand\andre[1]{{\color{magenta}[Andr{\'{e}}: #1]}}
  \newcommand\ian[1]{{\color{orange}[Ian: #1]}}
  \newcommand\ben[1]{{\color{blue}[Ben: #1]}}
  \newcommand\CWL[1]{{\color{brown}[CWL: #1]}}
\fi

\newif\ifrevise
\ifrevise
   \newcommand\change[1]{{\color{red}#1}}
\else
    \newcommand\change[1]{#1}
\fi

\newcommand\forceunits{\emph{E}$_\mathrm{H}$/\emph{a}$_\mathrm{B}$}

\newenvironment{sciabstract}{%
\begin{quote} \bf}
{\end{quote}}

\title{Accelerating Multiscale Electronic Stopping Power Predictions by 10 Million Times with a Combination of Time-Dependent Density Functional
Theory and Machine Learning}

\author
{Logan Ward,${}^{1\ast}$ Ben Blaiszik,${}^{1,3}$ Cheng-Wei Lee,${}^{2}$ Troy Martin${}^{3}$,\\ Ian Foster,${}^{1,3}$ André Schleife${}^{2\ast}$\\
\\
\normalsize{${}^{1}$Data Science and Learning Division, Argonne National Laboratory,}\\
\normalsize{Lemont, IL 60437, USA}\\
\normalsize{${}^{2}$Department of Materials Science and Engineering, University of Illinois at Urbana-Champaign}\\
\normalsize{Urbana, IL 61801, USA}\\
\normalsize{${}^{3}$Department of Computer Science, The University of Chicago}\\
\normalsize{Chicago, IL 60637, USA}\\
\\
\normalsize{$^\ast$To whom correspondence should be addressed; E-mail:  lward@anl.gov, schleife@illinois.edu}
}

\date{}

\begin{document} 


\baselineskip24pt


\maketitle 



\begin{sciabstract}
Knowing the rate at which particle radiation releases energy in a material, the ``stopping power,'' is key to designing nuclear reactors, medical treatments, semiconductor and quantum materials, and many other technologies.
While the nuclear contribution to stopping power, i.e., elastic scattering between atoms, is well understood in the literature, the route for gathering data on the electronic contribution has for decades remained costly and reliant on many simplifying assumptions, including that materials are isotropic. 
We establish a method that combines time-dependent density functional theory (TDDFT) and machine learning to reduce the time to assess new materials to hours on a supercomputer and provides valuable data on how atomic details influence electronic stopping. 
Our approach uses TDDFT to compute the electronic stopping 
from first principles in several directions and 
then machine learning to interpolate to other directions \change{at a cost of 10 million times fewer core-hours}.
We demonstrate the combined approach in a study of proton irradiation in aluminum and employ it to predict how the depth of maximum energy deposition, the ``Bragg Peak,'' varies depending on incident angle---a quantity otherwise inaccessible to modelers and far outside the scales of quantum mechanical simulations.
The lack of any experimental information requirement makes our method applicable to most materials, and its speed makes it a prime candidate for enabling quantum-to-continuum models of radiation damage.
The prospect of reusing valuable TDDFT data for training the model make our approach appealing for applications in the age of materials data science.
\end{sciabstract}


\section*{Introduction}


Particle radiation plays critical roles in modern society, including fabricating semiconductor electronics, characterizing materials and devices, cancer therapy, damage in nuclear reactors, and many others.
Advancing the application of particle radiation in these fields by maximizing and focusing their benefit and minimizing their detrimental impact requires precise control on microscopic length scales. Achieving such control relies critically on detailed fundamental understanding of how energetic particles interact with target materials.
Such understanding has been built for more than one hundred years \cite{bragg1905xxxix,bohr1913ii} through thorough experiments and sophisticated theoretical or computational models.
However, radiation experiments have high cost and low throughput, which ultimately can be attributed to high safety margins for experiments involving ionizing radiation.
These factors have motivated the development of intricate models that support experiments and eventually allow for predictions of the radiation-matter interaction and its consequences.

The high kinetic energies of ion beams and the energy-dependent response of the target material render radiation damage and the stopping power of the material, a friction-like, velocity-dependent force acting on radiation particles, inherently multi-scale problems.
Typically, one separation of scales is achieved by distinguishing the early stages of the process, where the projectile ion predominantly interacts with the electronic system of the target, from nuclear stopping, which happens only after the projectile slowed down significantly.
To deal with the large length and time scale aspect of radiation damage, the predominant mode for predictions relies on models around the binary collision approximation, parameterized by electronic and nuclear-stopping data collected, e.g., in the venerable Stopping and Range of Ions in Matter (SRIM) databases~\cite{ziegler2010srim,stoller2013use} as well as PSTAR/ASTAR\cite{Berger_NIST_report}, MSTAR\cite{Paul_ADNDT_2003}, DPass\cite{Schinner_NIMPRB_2019}, and Casp\cite{Schiwietz_NIMPRB_2012}.
Engineers can use these to predict the stopping distance of radiation or the amount of energy transferred to the target material, but within limits.
Electronic stopping powers tabulated, e.g., by SRIM for elemental compounds are based on Bethe-Bloch theory, with additional corrections to account for projectile charge state and relativistic effects, and are fitted to limited amounts of available experimental data \cite{ziegler2010srim,wittmaack2016misconceptions}.
Even fewer experimental results are available for multi-element compounds and Bragg’s rule of stopping power additivity is commonly used to close this gap \cite{KANG_CMS_2019}.
Specifically, such SRIM electronic stopping tables are generated by linear combination of their constituent elements, sometimes with a scaling factor if experimental results are known---an approach that works decently for high-kinetic energy (KE) projectiles but is less reliable in the low KE regime \cite{KANG_CMS_2019} (see Fig.~\ref{fig:introduction}).
In addition, atomic information like crystal structure and lattice orientation is not included in the electronic stopping tables.
These details, along with the target material's quantum-mechnical electronic structure, are critical in the low KE regime as well as for channeling projectiles. 
In short, SRIM has been a successful engineering tool for decades but its known faults leave much to be desired, especially for emerging modern applications, such as semiconductor qubits, electron microscopy, or proton cancer therapy.

On a microscopic level, nuclear stopping, i.e., elastic scattering between energetic particles and atoms in the target material, can be accurately modeled by Newtonian mechanics and has been well-studied in the past.
Molecular dynamics simulations of radiation damage have been performed for several decades \cite{Guinan:1981} and provide detailed and accurate understanding of defect cascades over large spatial and temporal scales of up to 100 nm and one nanosecond.
On the other hand, energetic projectiles with $v\approx v_{\mathrm{B}}$, i.e., kinetic energies of about 25 keV for protons (see Fig.~1),
also cause electronic excitations via inelastic scattering, i.e., electronic stopping. 
Quantum mechanics is needed to describe these interactions accurately, but first-principles methods remain costly for system sizes of technological interest that usually contain $\gg$ 1000 atoms.
Hence, in this work we focus on the electronic stopping contribution and how to address the large computational cost of computing it.

Before exploring the reduction of the computational cost, we introduce time-dependent density functional theory (TDDFT)~\cite{gross1990time} as an accurate ab initio route to predicting the electronic contribution to stopping power.
It is a practical formulation of the time-dependent Schr\"{o}dinger equation and  its real-time implementation (RT-TDDFT) has been shown 
in the past years to accurately predict electronic stopping of different particle radiations in diverse  material systems, including metals, semiconductors, insulators, and molecules (see Ref.\ \cite{Correa:2018} and references therein).
RT-TDDFT in principle can accurately address all the limitations of SRIM for electronic stopping but is currently limited to systems with at most a few hundred of atoms due to its computational cost. 
Efforts to incorporate RT-TDDFT results for electronic stopping into force-field-based molecular dynamics simulations of radiation damage have been accomplished, e.g., for Si \cite{Lee:2020}, but are limited to low projectile velocity ($v \ll v_{\mathrm{B}}$).
Full incorporation of electronic stopping data from RT-TDDFT into multi-scale simulations remains challenging, but at the same time bears the promise of extending SRIM and related approaches to new materials with unprecedented accuracy and predictive power.


Across the sciences, machine learning (ML) methods are increasingly opening the door for new scientific computing capabilities by reducing computational cost.
For example, machine-learned interatomic potentials have been used to increase the limits in system size and time scale in simulations of atomic-scale phenomena\cite{mishin2021mlipreview,vanderGiessen2020multiscale}, and
machine-learned models of atmospheric dynamics have provided a path for increasing the accuracy and reducing the cost of modeling earth systems\cite{maulik2020geoml, rasp2018geoml}.
The challenge behind all of these successes is that building a surrogate model to supplement physics-based approaches is non-trivial.
Scientists must contend with limited and potentially biased data\cite{roberts2021covidml,Guo:2022}, ensure that their model predictions reflect laws of physics \cite{anderson2019covariantnn, rupp2015nutshell}, and leverage existing knowledge to develop highly accurate models.
Building a successful surrogate therefore requires careful consideration of the problem and how to validate the model before one can use it to perform new science.

Here, we demonstrate how the rich data produced by RT-TDDFT simulations can be combined with ML to expand greatly the ability to predict the electronic stopping power of materials, enabling dramatically improved speed while maintaining the accuracy of first principles results.
We perform a detailed study of a ML model that predicts the velocity-dependent electronic stopping of a proton traveling through face-centered cubic (FCC) aluminum.
Our ML model accurately predicts electronic stopping in crystal directions absent from the training data at rates 10\textsuperscript{7} times faster than TD-DFT. 
Critically, our work required no new DFT calculations: rather, as we describe below, it reuses 10 trajectories computed by Schleife et al.\ in 2015~\cite{schleife2015tddftstopping} and now available~\cite{2015data} in the Materials Data Facility (MDF)~\cite{blaiszik2016materials}.
Thus this work opens an exciting ab initio route for predicting electronic stopping fast enough to interface with molecular dynamics simulations, where it was recently shown that electronic stopping is an important factor determining emerging damage cascades\cite{Lee:2020,Jarrin:2021}.
We anticipate that this work paves the way towards a significantly improved understanding of radiation damage in a large number of materials.
The work also suggests new avenues for data reuse in materials science, which---outside narrow areas such as materials property prediction~\cite{jain2016research}---remains relatively rare\cite{himanen2019data,suhr2020search,brinson2022fair}. 

\section*{Results}

\begin{figure}
    \centering
    \includegraphics{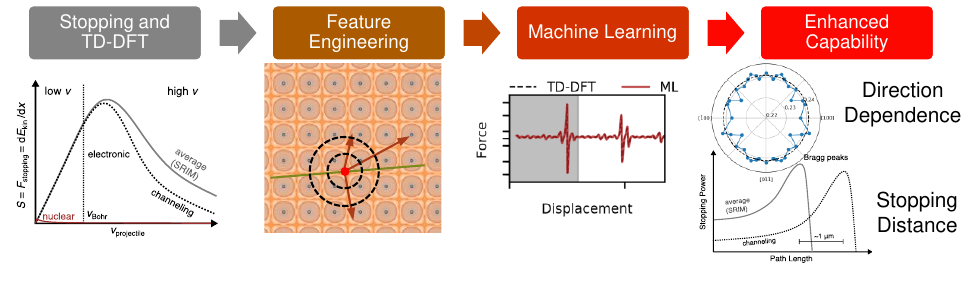}
    \caption{We extend the capability of Time-Dependent Density Functional Theory (TD-DFT) by building a surrogate model that predicts the drag force on a projectile based on its atomic environment and local electron density. The machine learning surrogate operates 10\textsuperscript{7}$\times$ faster than TD-DFT which enables first-principles assessments of direction-dependent stopping forces and full computations of stopping distances.}
    \label{fig:introduction}
\end{figure}

Stopping power is defined as the energy loss per distance traveled by the projectile ion\cite{schleife2015tddftstopping}, which has the unit of force and is proportional to energy loss rate.
A projectile traveling through a target material at velocities on the order of the Bohr velocity, i.e., the velocity of an electron in the first Bohr orbital of a hydrogen atom, experiences predominately electronic stopping (see Fig.~\ref{fig:introduction}).
We model electronic stopping by using real-time time-dependent density functional theory (RT-TDDFT) to simulate a hydrogen atom traveling through FCC aluminum at constant velocities, since the femto-second time scale of interest is too short for projectile velocity to change significantly, less than $<$ 0.4\% during the entire simulation\cite{schleife2015tddftstopping}. 
Electronic stopping can then be calculated from RT-TDDFT results by tracking either the energy change of the electronic system of the target material as the projectile travels through it or the force that the projectile experiences against its traveling direction \cite{schleife2015tddftstopping}.
Schleife et al.\ previously demonstrated that both approaches result in the same electronic stopping for proton-irradiated Aluminum\cite{schleife2015tddftstopping}.
Used this way, RT-TDDFT provides exclusively electronic stopping, since the projectile cannot transfer any momentum to the fixed target atoms and all ionic stopping contributions average to zero.
In the following sections, we detail a process for training a machine learned surrogate model and then describe how we validated our approach by predicting electronic stopping power in Aluminum.
We conclude with an example of how, due to their extremely reduced computational cost relative to RT-TDDFT, our new surrogate models provide previously inaccessible capabilities, such as computing direction-dependent Bragg peaks with first-principles accuracy.
We emphasize that by training on RT-TDDFT data, our approach can be straightforwardly extended to any material.

\subsection*{Developing a machine-learning surrogate for stopping power calculations}

The first step in creating our ML surrogate model for RT-TDDFT electronic stopping predictions is to identify what inputs and outputs are required.
In a RT-TDDFT calculation, the input to each time step comprises of the Kohn-Sham (KS) electronic states $\psi(\mathbf{r},t)$ and the positions and velocities of all ions at that time $t$.
Using this, the TDKS equations, Eq.~\eqref{eq:tdks}, are explicitly propagated in real time \cite{schleife:2012,draeger:2017} by $\Delta t$, producing $\psi(\mathbf{r},t+\Delta t)$ and the corresponding electronic total energy $E(t+\Delta t)$.
Simultaneously, the equation of motion for the ions is also propagated in real time by using Hellman-Feynman forces that are computed from the time-dependent electron density $n(t)=\sum_i^N{\left|\psi(\mathbf{r},t)\right|^2}$, providing new atomic positions and velocities \cite{Marx:2009}.
The ability of RT-TDDFT to model the state of quantum-mechanical electrons and classical atoms over time gives it a rich suite of capabilities, of which we require only a subset to predict electronic stopping power.

We choose to emulate enough of the RT-TDDFT calculations to compute direction dependent stopping, but not so much as to require quantum-mechanical computations when using the surrogate.
In this work we use position and velocity of the projectile and the ground state electron density as input, and the stopping force is the output.
Our choice avoids modeling the explicit time evolution of the electronic wave functions, which is a long-standing problem with machine learning for excited states \cite{westermayr2020mlexcited}.
Using position and velocity to determine the projectile's present state allows making predictions of the stopping force at any projectile position without needing to first evolve the system to account for projectile history.
By excluding any explicit history dependence, we assume that the projectile's initial conditions (e.g., initial charge state) no longer affect the stopping force, which is the same assumption used when computing stopping force by using RT-TDDFT, where long enough simulations are performed to ensure independence of initial conditions by means of charge equilibration.
We also discount any natural variations in the stopping power between crystallographically-identical projectile states that occur due to the time dependence in the charge localized on the projectile, which is averaged over when determining stopping power in RT-TDDFT over multiple unit cells.
Our choice of inputs (position, velocity, charge density) is comparable to parameters used in physically derived models of electronic stopping, such as the Lindhard\cite{Lindhard_1964} and the Firsov\cite{firsov1959qualitative} models, which are functions of velocity and local charge density (Lindhard model) or impact parameter (Firsov model).

Next, we determine how to transform these inputs into a set of variables (``features'') that would serve as useful inputs to an ML model.
As well articulated by Faber et al.\cite{faber2015intj}, features must succinctly capture the essential physics driving the stopping force and must be quick to compute.
One of our features is the force resulting from the repulsion between the projectile and the atomic nuclei, which we compute explicitly by calculating the Coulombic force between positive charges by using Ewald summations \cite{ewald1921}.
Lacking inexpensive theoretical models for the interaction between projectile and electrons, we approximate the effects of interactions with electrons implicitly using features of the local electron density and the distribution of atoms around the projectile.
From the ground-state DFT charge density we selected several points at fixed distances in front of and behind the projectile as features.
The local density of atoms is captured by using the AGNI fingerprints of Botu et al.\cite{botu2017machine}, which describe the density of atoms at specific distances away from the projectile along specific directions (Eq.~\ref{eq:agni}).
We chose to use AGNI fingerprints along the projectile's direction of travel.
In summary, our full set of features describes the state of the projectile by using the ion-ion Coulomb repulsion, the local distribution of atoms, and the ground-state electron density.
Full details of the features are available in the Methods section, and we analyze and discuss their relative importance later.

Our last task is to use ML to map the variables used to describe the state of the projectile to the force acting on it.
There are many regression algorithms, but only a few possess the characteristics needed for our task, i.e., continuous derivatives with respect to projectile position and ability to train accurate models with a moderate amount of training data, as we estimate having access to $\approx 10^5$ training examples.
We therefore limit our search to different types of linear regression methods and neural networks and demonstrate how to identify the ML algorithms with the highest generalizability in the following sections.

\subsection*{Re-using data from previous studies}

We selected the electronic stopping of a proton traveling through face-centered cubic aluminum as a case study.
Schleife et al.\ have previously demonstrated that RT-TDDFT can reproduce experimental stopping powers for this case \cite{schleife2015tddftstopping}.
The data produced from that validation are extensive, including stopping powers for multiple projectile directions through the material at many velocities---a total of \num{58969} pairs of projectile positions and stopping forces.
The availability of those legacy data meant that we could train our ML models without the need for any new RT-TDDFT calculations \cite{2015data}.
We denote this data as the Schleife 2015 dataset.

The Schleife 2015 dataset includes a total of 10 trajectories, each representing a proton traveling in a certain direction at a constant velocity through bulk aluminum.
The projectile speeds vary from 0.5 to 4 atomic units (a.u.) of velocity, with 1 a.u.\ corresponding to roughly 25 keV of kinetic energy for proton projectiles.
Thus the data span both the linear-response region where electronic stopping power increases linearly with projectile speed and the high-velocity regime where the stopping power decreases with speed.
The data also include two directions: a ``hyper channel,''
$\left<\textrm{100}\right>$,
direction that passes maximally distant from any atom in the material and a randomly-selected direction that passes near some atoms and samples a much larger range of distances between projectile and aluminum atoms.
The variety in speeds and projectile environments makes the dataset ideal for characterizing the stopping power of a material and training a generalizable ML model.
We note that this same variety is needed to predict electronic stopping comprehensively from first principles for any material by using RT-TDDFT simulations \cite{schleife2015tddftstopping}.

\subsection*{Training a model to predict forces}

\begin{figure}
    \centering
    \includegraphics[width=\textwidth,clip]{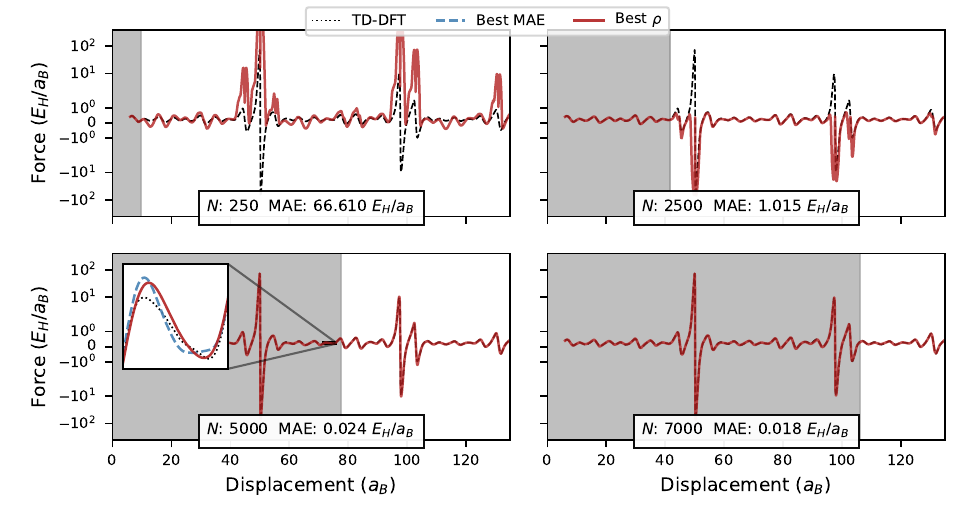}
    \caption{
    \label{fig:time_evolution}
    Force acting on a proton projectile passing through aluminum as calculated by RT-TDDFT (black, dotted line) and predicted by a ML model (red, solid line) trained on progressively larger training sets.
    The data from timesteps in the gray region were used to train the ML model in each frame.
    The mean absolute error (MAE) is measured across the entire trajectory.
    The inset in the bottom left panel shows the performance of a model selected based on MAE (blue, dashed line), Bayesian linear regression without feature selection, compared to our model which was chosen based on Spearman's correlation coefficient $\rho$, Bayesian linear regression with LASSO-based feature selection.
    Predictions from the model selected by using MAE are less smooth than either RT-TDDFT or the model based on $\rho$.
    }
\end{figure}


The first problem we considered was the prediction of stopping power in multiple directions at a \emph{single} projectile speed.
We selected an intermediate projectile speed from our available training data, 1.0 atomic units, as a starting point. 
The goal is to train a model that can both predict forces on the projectile at future timesteps of the same trajectory and generalize to trajectories not included in the training set. 
Our initial task was to identify an appropriate ML algorithm that can achieve optimal predictive performance.
The ML model must also have continuous derivatives and have training times less than several hours with up to 10\textsuperscript{5} training examples.
We enumerated a total of eight linear-regression-based learning algorithms that vary based on whether they include polynomial features and regularization approaches (e.g., LASSO, Bayesian Ridge, recursive feature elimination).
A full list and software needed to recreate our models is available along with the data published with this publication \cite{ward2023sidata}.

We selected an appropriate algorithm by training each on the first half of the data for the random trajectory, then measuring performance on the entire random trajectory.
The random trajectory samples more atomic environments than $\left<\textrm{100}\right>$ does and has a larger number of samples, which makes it a better source of training data.
We measured performance across the entire random trajectory to capture both fitness to the training data and generalizability.
We use the mean absolute error (MAE) of the predictions and Spearman’s correlation coefficient $\rho$, which measures the quality of the ranking from lowest to highest force, to evaluate the performance of the models.
We provide full performance data for the different ML algorithms in the Supplementary Information. 

Models with the best Spearman's correlation coefficient have better behavior than those with the lowest MAE.
As shown in the inset of Fig.~\ref{fig:time_evolution}, the model with the best MAE, Bayesian ridge regression without feature selection, has slight deviations in the predicted forces compared to the training data.
The problem stems from the fact that the MAE takes the scale of the data into account;
therefore, models selected by using this metric may be biased towards those that fit the high-force regions better.
Fitting to larger forces better is a problem because, as shown in Fig.~\ref{fig:time_evolution}, the force experienced by the projectile varies over several orders of magnitude.
Fitting to only the high-force regions is a concern because the low-force regions are more prevalent and thus more important for stopping (note how the projectile still experiences stopping forces in the channel) and the accuracy of the training data in the high-force regions is limited due to the use of a pseudopotential in our simulations.
The model with the best Spearman's correlation coefficient lacks these deviations and achieves a better accuracy on regions where the magnitude of the force is small (MAE: 0.0142~\forceunits{}) than the model with the best overall MAE (MAE: 0.0174~\forceunits{}).
Consequently, we selected a Bayesian ridge regression model that uses a reduced subset of polynomial features identified as important via LASSO;
this lacks the deviations while also maintaining close agreement with RT-TDDFT in the low-force regions.


\begin{figure}
\centering
\includegraphics{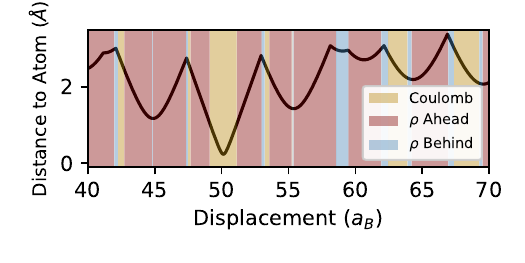}
\caption{
\label{fig:features}
The group of features that are most influential in predicting the stopping force as a function of projectile displacement in a random trajectory with $\left|v\right|$ = 1.0 a.u.
Features are grouped into those that include Coulomb repulsion between the projectile and nearby nuclei, and those related to the charge density ($\rho$) ahead of or behind the projectile.
The color of the background indicates which group of features is most important over a certain range of displacements.
The distance to the closest atom is shown as a black line.
}
\end{figure}

The features of our ML model selected by using LASSO correspond to a variety of physical effects underlying electronic stopping (see Fig.\ \ref{fig:features}). 
Only 25 of the 189 features generated via polynomial expansion  are selected and they fall into three categories:
features that include the ion-ion repulsion, features based on the charge density ahead of the projectile, and those based on the charge density behind.
The values of the features do not vary independently, so we assess the importance by computing their influence on predicted force at different projectile positions.
The Coulomb repulsion is particularly important as the projectile passes close to an aluminum atom, which is clearly visible near a displacement of 50 $a_\mathrm{B}$ in Fig.~\ref{fig:features}.
Features related to the charge density ahead of the projectile are important most often---presenting a mechanism of a projectile ``pushing'' through a sea of electrons.
The charge density behind is only important in small sections of the trajectory, which shows it is of limited importance but could be indicative of a projectile's history having direct influence on stopping force.
In short, the feature importance is consistent with known physics of stopping forces (e.g., ion-ion repulsion becoming important at short ranges) but also captures effects yet to be incorporated into physics models.

We further characterize our Bayesian ridge regression model by evaluating the performance as more training data is provided. 
As shown in Fig.~\ref{fig:time_evolution}, the model can accurately reproduce the force in all low stopping power regions when provided with only the first quarter of the trajectory.
However, in this case the model has yet to be trained on regions of large forces, by using training examples corresponding to near passes of the projectile to an atomic core, and we note the model performs poorly when predicting the stopping force for those regions.
The model accurately recreates the entire training trajectory once 50\% of the data are included in the training set, which also includes data from the low- and high-force regions of the trajectory.
The model then accurately reproduces the stopping force in the major peak near a displacement of 100~$a_B$, without those data being included in the training set.
Increasing the amount of training data to 75\% of the trajectory continues to lower the error of the model, but with diminishing returns. At this point, the features identified by the LASSO feature selection technique and coefficients determined by Bayesian ridge regression are effectively converged.

\begin{figure}
    \centering
    \includegraphics[width=0.5\textwidth]{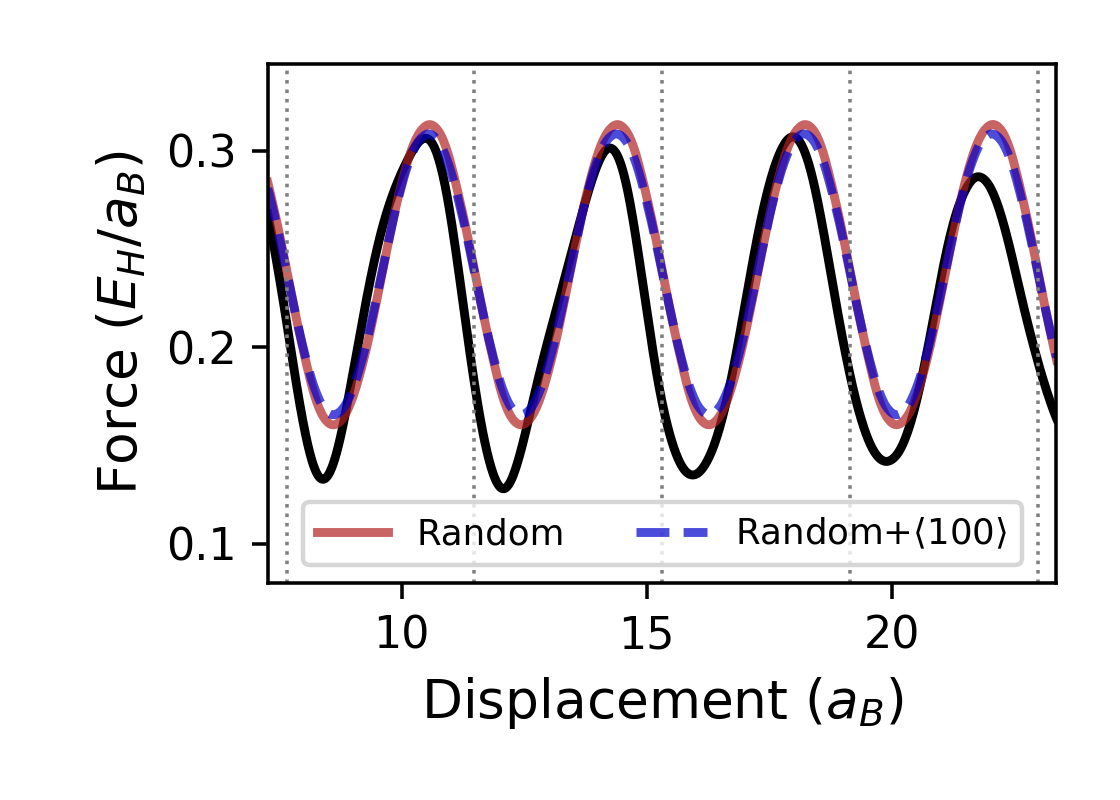}
    \caption{
    \label{fig:channel_pred}
    Predicted stopping force acting on a proton moving along the $\left<\textrm{100}\right>$ channel of FCC Aluminum as computed with RT-TDDFT (black, \change{solid}), as predicted by an ML model trained by using RT-TDDFT data from a randomly-oriented trajectory (red, solid), and an ML model trained on both the channel and random data (blue, dashed).
    Vertical gray lines indicate the boundaries of the unit cells.
    We show the second and third unit cells of the six available in our training data.
    }
\end{figure}

Next, we validated this model by predicting the stopping power in different crystallographic directions, starting with the $\left<\textrm{100}\right>$ channel.
As shown in Fig.~\ref{fig:channel_pred}, the model correctly captures many key behaviors, such as the sinusoidal character of the stopping force with a periodicity of half a unit cell length.
The model has approximately the correct amplitude but overestimates the average force by 0.023~\forceunits{}, an 11\% error with respect to the average stopping power in the channel.
An error of 0.023~\forceunits{} is within acceptable bounds as it is smaller than the differences we are seeking to quantify; stopping power changes by up to 0.06~\forceunits{} (see Fig.~\ref{fig:stopping-power-val}) depending on projectile direction and varies more than 0.1~\forceunits{} depending on velocity (see Fig.~\ref{fig:multi-velocity}).
Our ML model also predicts the position and magnitude of the maximum force well, just before the projectile passes nearest to an atom, e.g., at the unit cell boundary.
It is slightly out-of-phase and Fig.\ \ref{fig:channel_pred} shows that the model prediction is off by 0.19 $a_\mathrm{B}$ (6.8\%) in location and 0.014 \forceunits{} (4.6\%) in value.

To put the error of the ML model into context, we compared it to the best possible agreement with RT-TDDFT achievable on the $\left<100\right>$ channel. 
Our ML model predicts the same stopping force for two projectiles with the same velocity at symmetrically-equivalent positions in a lattice, but this is not true in our TD-DFT simulations and implies a bound on the accuracy of our predictions.
The RT-TDDFT forces can be different both due to 
remaining transient effects from inserting the projectile into the lattice (the two projectiles may differ in their time since simulation start),\cite{schleife2015tddftstopping} and effects due to the finite size of the simulation cell (projectiles may interact with their periodic image in varying amounts).
Our ML model accounts for none of these effects, which makes the variation between identical projectile positions and velocities in RT-TDDFT a good comparison to measure predictive accuracy.
We assessed an average deviation of 0.0063~\forceunits{} between the force at equivalent positions in the four unit cells from our training data, which represents the magnitude of these effects ignored by our model.
The MAE of 0.025~\forceunits{} determined for our model is four times larger than this uncertainty figure.
A corresponding upper-bound on accuracy is assuming that force is independent of position, which would result in an MAE of 0.050~\forceunits{}, i.e., two times larger than our model.
Considering both the qualitative and quantitative agreement of the model to the channel direction, we conclude that our ML strategy can interpolate stopping forces on trajectories outside of our training set.


\subsection*{Accurately assessing direction dependence}

\begin{figure}
\centering
\includegraphics{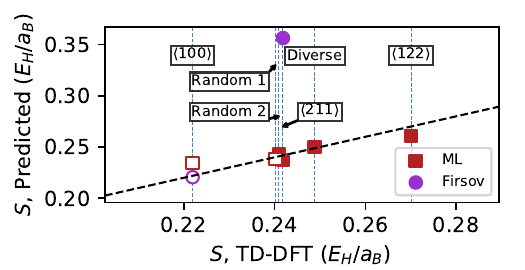}
\caption{\label{fig:stopping-power-val}
Average stopping powers at $v=$ 1.0 a.u.\ computed by RT-TDDFT for six different trajectories in FCC aluminium compared to an ML model (red squares) and the Firsov model (purple circles).
Both ML and Firsov models are parameterized by using RT-TDDFT data from the trajectories marked with white fill.
Trajectories used to validate model performance are shown with colored fill.
Points which lie closer to the black, dashed line ($y=x$) are better predictions.
The two random trajectories start at the same position as the $\left<100\right>$ trajectory but travel on different directions, 
and the diverse trajectory was selected to sample as many atomic environments as possible.}
\end{figure}

The stopping power of a material varies strongly depending on the direction along which a projectile travels.
For example, some directions result in a projectile interacting with the dense clouds of electrons around atoms more frequently 
than others and, therefore, loosing energy faster.
The exact differences in stopping power are important for designing materials and currently only accessible via RT-TDDFT.
We explore here if our ML models offer an~alternative.

We performed additional RT-TDDFT calculations to assess direction dependencies.
These include a second randomly selected direction, a direction with a moderate stopping power ($\left<\textrm{211}\right>$), a  direction where the ML model predicts an especially low stopping power ($\left<\textrm{122}\right>$), and a direction that passes through a diverse set of atomic environments.
The diverse channel was determined by finding a trajectory which maximizes the diversity of environments observed by the projectile over a certain distance (see SI for details) and is the same length as the random directory.
A large diversity of training environments provides a challenging test set and, as illustrated by recent work of Kononov et al.\cite{kononov2023trajopt}, a potential route for reducing the data requirements of RT-TDDFT.

We trained our ML model on the random and $\left<\textrm{100}\right>$ channel trajectories to ensure the best accuracy, and then used it both to predict the forces across each trajectory.
The resulting stopping power predictions in Fig.~\ref{fig:stopping-power-val} illustrate strong qualitative accuracy, but with notable drops in accuracy for the most atypical channels.
The ML model overestimates the stopping power of the trajectory with the lowest stopping power ($\left<\textrm{100}\right>$ channel) and underestimates the high stopping power for the diverse channel.
Low predictive power for extreme values is a common issue with ML models because forecasting in the rare regimes that cause them, by their nature, require a greater degree of extrapolation.
Regardless, the error of 5\% for these extreme points is comparable to the error between RT-TDDFT and experiment of $\sim$5\%\cite{schleife2015tddftstopping}
which indicates that the limiting factor on the accuracy of using this model will be between RT-TDDFT and experiment instead of between ML and RT-TDDFT.

Our model also compares favorably to the Firsov model, a well-established analytical method for predicting the electronic stopping power of heavy ions in materials in the low KE regime~\cite{firsov1959qualitative,Sillanpaa_PRB_2000}.
The Firsov model is not perfectly suited for our applications; being designed for heavy ions (we study protons), channel trajectories (we consider closer passes to atoms), and the linear scaling regime (we explore velocities higher than that range). 
That said, the Firsov model is the best-available theory because it depends on projectile position in the lattice and we can make a fairer comparison to ML by parameterizing the Firsov model by using data from one channel without close passes near atoms ($\left<100\right>$) and test against the other in our dataset ($\left<211\right>$). 
The Firsov model overestimates the stopping force of the second channel by 50\% (RT-TDDFT: 0.24, Firsov: 0.36~\forceunits{}), which is much worse than the 2\% underestimation from machine learning (ML: 0.237~\forceunits{}).
Machine learning not only predicts stopping power more accurately in channels where Firsov is applicable, but can also extend to trajectories where Firsov cannot.


\subsection*{Million-fold acceleration of stopping power prediction}

Stopping power is computed from RT-TDDFT by averaging stopping force over a trajectory.
RT-TDDFT calculations rely on computing the stopping force at each timestep sequentially---a process that requires 2.1~core-hours per timestep for the simulations used in our study\change{, which ran QBox\cite{gygi2008qbox} on 768 Intel Xeon X5660 cores of the Sierra supercomputer.}
The number of timesteps varies with the length of the trajectory before it repeats over the same environment due to crystalline symmetry.
The $\left<\textrm{100}\right>$ channel is the shortest trajectory in FCC aluminium and still requires around 2000 timesteps for stopping power calculation at a projectile speed of 1.0 a.u.\ which totals a significant 4028 core-hours.
Here, we establish by how much we can reduce this computational cost.

Simply replacing RT-TDDFT with our ML model and repeating the evaluation of each of the 2000 timesteps needed to compute the stopping power in the $\left<\textrm{100}\right>$ channel requires only 13~core-s on a \change{AMD Ryzen 9 5900X}, i.e., a million-fold cost reduction from TD-DFT, and this can be further improved.
We further accelerate the computation of stopping power with adaptive integration techniques, which sample a function more densely in regions where it varies more than in regions where the function is constant.
We use adaptive Gaussian quadrature, as implemented in QUADPACK\cite{piessens1983quadpack}, to pick the smallest number of points required to achieve a stopping power accurate to 10\textsuperscript{-3} \forceunits{}.
Adaptive quadrature evaluates forces at positions out of sequence with time departs, which departs strongly from the sequential timestepping of current RT-TDDFT algorithms but is easy with how our ML algorithm is designed.
Adaptive quadrature speeds the calculation of the channel stopping force by another 30 times to a total of a factor of 3 $\times$ 10\textsuperscript{7} compared to the RT-TDDFT simulation. 

\subsection*{Accounting for velocity-dependence of electronic stopping}

Electronic stopping power of a projectile in target materials has a complex dependency not just on trajectory, but also on velocity.
It is well-established that electronic stopping power in metals generally increases linearly as a function of velocity at low velocity ($v < 1$~a.u. for Aluminum), up to a maximum of the stopping power at around $v \approx 1.5$~a.u., after which it decays back to zero at increasing velocities (see Fig.~\ref{fig:introduction}).
While precise prediction of the rates of these effects yet eludes theory, recent RT-TDDFT studies provided more insight, e.g., on the dependency of electronic stopping on the electronic structure of the target material, the effective charge state of the projectile, and its spatial proximity to the atoms of the target material.
Band-structure effects cause deviations from the linear behavior at low velocity \cite{Quashie_PRB_2016}. 
At high projectile velocity, core electrons contribute to electronic stopping for projectiles on off-channeling trajectories \cite{schleife2015tddftstopping,Lee:2020,Yao_PRL_2019,Ullah_PRL_2018}.
In addition, the equilibrium charge state itself, e.g., of an initially highly charged ion projectile, depends on the proximity to target atoms \cite{Lee:2020}.
This complex interplay of dynamic projectile charge, trajectory, and electronic structure of the target make predicting the velocity dependence of electronic stopping challenging.
While first-principles RT-TDDFT simulations in principle capture this complexity, no analytic model can achieve this and existing approximations, such as the linear model used by Firsov, are not valid across the entire velocity range.
Hence, after establishing in the previous section that our model is capable of describing the physics of electronic stopping at a single velocity, we now explore how ML can learn the subtlety of velocity dependence.

Our first attempt was to add velocity as an input feature to the model and then use the same polynomial regression algorithm as for the single-velocity model.
We trained the model on the data from the random trajectory with projectiles traveling at speeds of 0.5, 1.0, 2.0, 3.0, and 4.0 a.u.
However, the linear regression model fails to adequately fit the training set when we include more than one velocity in the training data.
Training on the random trajectory yields a mean absolute percent error (MAPE) of 16\% in predicting stopping power of the random channel at different velocities---far higher than the 0.4\% error in stopping power when trained on a single velocity. 
Without the ability to fit the training set accurately, we concluded that the linear models are too simple to capture the complexity of velocity-dependence.

Following our hypothesis of insufficient model complexity, we decided to use a neural network to fit the training data.
Neural networks fit the requirements for a ML model outlined earlier, provided we design our network architecture appropriately, yet allow us to express more complex functions.
The model will have continuous derivatives, if we choose activation layers with continuous derivatives.
After some experimentation with different network designs, we found that a simple multi-layer perceptron with six hidden layers (2015 parameters in total) and exponential linear unit (ELU) activation functions\cite{clevert2015elu} both achieves strong fitness to the training set and generalizes to our test sets.
Our neural network model achieves a MAPE of only 0.25\% in predicting the stopping powers of the random direction when trained on the random direction, which is 64 times lower than the polynomial regression model in the same test case.
The increased complexity of a neural network may even have advantages for our earlier, single velocity problem but at the cost of lower interpretability and speed.

The neural network model interpolates to other directions just as well as the single-velocity, polynomial regression model.
We trained the neural network on the data from the random trajectory and $\left<\textrm{100}\right>$ channel directions at all five velocities available in the training set and evaluated its performance on the six directions available at $\left|v\right|$ = 1.0. 
The MAPE of the stopping powers predicted by using the neural network is 1.9\%, which is equivalent to the MAPE of 2.1\% for the polynomial regression model trained on only the $\left|v\right|$ = 1.0 data.
From this, we conclude that the neural network is an appropriate choice for models that account for changes in trajectory and speed of the projectile.

\begin{figure}
\centering
\includegraphics{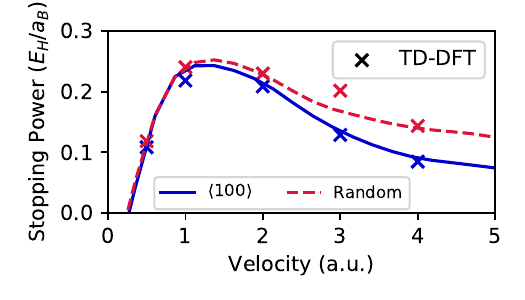}
\caption{\label{fig:multi-velocity}
Stopping power as a function of velocity magnitude predicted by a neural network model for the $\left<\textrm{100}\right>$ hyper-channel (blue, solid) and a randomly-selected off-channel direction (red, dashed).
The network was trained on the stopping force as a function of position and velocity for the hyper-channel and random direction at five speeds between 0.5 and 4~a.u. 
The stopping power for these 10 TD-DFT trajectories used for training are shown as Xs.}
\end{figure}

Beyond quantitative accuracy on specific points in our test set, the neural network model also captures qualitative trends for how electronic stopping power is affected by velocity.
In Fig.~\ref{fig:multi-velocity}, we show the stopping power as a function of increasing speed for two different directions:
the $\left<\textrm{100}\right>$ hyper-channel and the random direction, which passes closer to atomic nuclei.
At low velocities, the stopping power for both the channel and random trajectory increases linearly as a function of velocity.
Further, electronic stopping for these two trajectories is similar at speeds below 1.5 a.u.
Projectiles traveling at these slow velocities are not energetic enough to excite core electrons---leading to small differences in stopping power between whether a projectile traveled near atomic nuclei or not.
The stopping power in the channel and random directions diverges increasingly as velocity increases past 1~a.u., which is expected because of the possibility of exciting core electrons in this velocity regime \cite{schleife2015tddftstopping}.
The direction dependence of our neural network model even outperforms the linear regression model created earlier.
The neural network model achieves a mean absolute percentage error (MAPE) of 1.9\% on the holdout set shown in Fig.~\ref{fig:stopping-power-val}, compared to 2.2\% for the linear regression.
We note that direction and velocity effects were not explicitly coded into the model but are properties of electronic stopping learned without human intervention;
the ML model was able to infer this behavior from the data.

The ML model deviates from expected behavior at velocities higher and lower than those in our training set, i.e., in the extrapolative regime.
For one, the stopping power should approach zero as velocity increases to very high velocities. 
The stopping powers predicted from our model instead become negative at extreme velocities ($v>$ 10~a.u.), which is unphysical.
The stopping power should also approach zero as velocity approaches zero, but our ML model shows a zero stopping power for finite velocities (see Fig.~\ref{fig:multi-velocity}).
Zero electronic stopping power at finite velocities is a feature of semiconductors and insulators, i.e.,  materials with a non-zero bandgap, as they require certain energies to excite electrons.
However, aluminum is a metal and has no band gap.
Our model is unable to make adequate predictions at these extreme velocities, as data in these regimes is absent from the training set and the model architecture was not designed to encode such limiting behavior.

In short, we establish both that our model is a suitable interpolator for predicting stopping power as a function of direction and speed, and also established limits of the applicability of the model.
Our model has excellent quantitative accuracy at predicting both the instantaneous force acting on projectiles along and stopping power of many different trajectories.
We also find that the model reflects known physical laws within the regions for which we have training data:
The stopping power increases linearly at low velocities and the effect of direction becomes larger with increased velocity---in both cases, as expected from theory and reflected in the training data.
However, the model fails to capture the fact that the difference between channel and off-channel should diminish with velocity, and that stopping is non-zero even at slow velocities.
Consequently, we recommend that these models only be used at intermediate velocities (0.5--4.0 a.u.); to improve accuracy in other regimes, we recommend either adding theory-based constraints to the model or training with more data in those velocity regimes.

\subsection*{Example application: Predicting Bragg Peak distribution}


Radiation particles release energy unequally through a solid and typically deposit the largest amounts of dose towards the end of their trajectory.
This localization of energy release is key for many technological applications of radiation, such as cancer treatments, ion implantation, and creation of optically active defects in semiconductor qubits\cite{Krasheninnikov_JAP_2010, Haume_CN_2016,Lee_NL_2019,Wolfowicz_NRM_2021}.
The position of maximum energy release is known as the ``Bragg peak.''
The current route for predicting the Bragg peak is based on SRIM, which---as noted previously---assumes that the stopping forces are independent of direction.
To go beyond SRIM by incorporating details of the crystal structure for nuclear stopping, two approaches were previously introduced.
The first one utilizes the same binary collision approximation as SRIM but applies it to crystalline target materials \cite{MARLOWE_code_2014,Crystal-trim_1994}.
The second approach utilizes molecular dynamics simulations to simultaneously address the limitations of the binary collision approximation in the low kinetic energy regime \cite{Sillanpaa_PRB_2000,hinks2018mdrange}.
However, these approaches generally utilize semi-empirical models for electronic stopping, which lack crystal-structure information.
A recent study, utilizes RT-TDDFT along with molecular dynamics simulations to provide a more general route for accessing structure- and direction-dependent stopping distances for not only nuclear but also electronic stopping, however, this comes at large computational costs \cite{sand2019tddftionranges}.
As a result, this work is limited to a specific channeling direction, $\left<\textrm{001}\right>$ in Tungsten. 
Our ML model provides a computationally efficient route to calculate electronic stopping under different impact parameters and velocities and thus can augment the combined approach of molecular dynamics and RT-TDDFT. 
For instance, focusing only on the contribution of electronic stopping, our ML model can predict how the Bragg peak's location changes with different crystal orientations and establish the distribution of how it varies across parallel trajectories.

\begin{figure}
\centering
\includegraphics{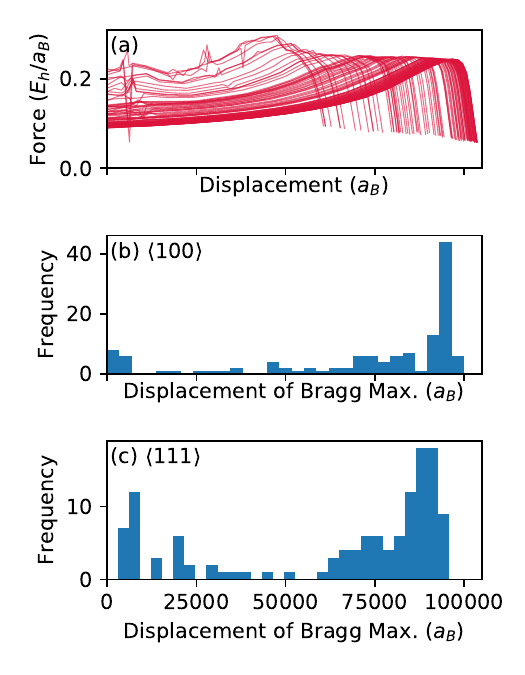}
\caption{
\label{fig:bragg}
Analysis of the stopping force as a function of distance for projectiles traveling along the same direction but with different starting positions.
(a) Stopping force as a function of displacement for projectiles traveling in the $\left<\textrm{100}\right>$ direction as predicted by our ML model.
(b,c) Distribution of the position of maximum stopping force for 128 trajectories in the (b) $\left<\textrm{100}\right>$  and (c) $\left<\textrm{111}\right>$  directions, as computed by our ML model.
The median stopping distance is 98.8 $\times$ 10\textsuperscript{3} $a_B$ for the $\left<\textrm{100}\right>$ direction compared to 87.0 $\times$ 10\textsuperscript{3} $a_B$ for the $\left<\textrm{111}\right>$ direction.
}
\end{figure}

We started by evaluating the Bragg peak location for projectiles travelling along the $\left<\textrm{100}\right>$ direction of aluminum.
Each projectile starts at a random position in the simulation cell with an initial speed of 4.0~a.u. and we propagate its position by using Newton's laws of motion until the velocity reaches 0.4~a.u.
The Bragg peak is the region on the trajectory where the force is the largest, which we determine from a smoothing spline fit to the force over distance (see Fig.~\ref{fig:bragg}).
We select 0.4~a.u.\ as a minimum velocity as we observe it to be significantly below the velocity at which the Bragg Peak occurs and because it is slower than any trajectory in our training set.
We do not update the positions of the surrounding atoms, which means our approach does not include any nuclear stopping. Nuclear stopping only matters once the projectile is slow enough to interact with atoms of the target by causing damage via displacing ions or, eventually, implanting inside the target material \cite{sand2019tddftionranges}.

We estimate the variation in the location of the Bragg peak by starting the projectiles from different positions within the aluminum unit cell and then computing the stopping force as a function of displacement for each trajectory.
We observe a Bragg peak in many of the $\left<\textrm{100}\right>$ trajectories (see Fig.~\ref{fig:bragg}a) and a large variation in the depth depending on the starting condition (see Fig.~\ref{fig:bragg}b).
The interquartile range of the stopping distance is 33 $\times$ 10\textsuperscript{3} $a_B$, which is 40\% of the median depth.
Producing this estimate required simulating a total of 0.1~ns of electronic stopping, which would have required 300M timesteps of RT-TDDFT and 700M core-hours of compute time if simulated directly (1.5 $\times$ 10\textsuperscript{3} tons CO\textsubscript{2})\footnote{Estimated by using the power per core of Vulcan, 5~W, and 2021 average US power generation, 0.855~lb/kWh}.

The variation in Bragg peak location is especially visible between projectiles that travel along different directions.
In particular, projectiles traveling in the $\left<\textrm{111}\right>$ direction illustrate both how the mean position of the Bragg peak changes but also how the variance can increase depending on direction (see Fig.~\ref{fig:bragg}c).
Projectiles traveling in the $\left<\textrm{111}\right>$ direction not only have Bragg peaks 14\% closer to the surface, but also disperse their energy over a wider area.
The difference between the 75th percentile and the maximum of the Bragg peak distance is four times greater in the $\left<\textrm{111}\right>$ direction than in the $\left<\textrm{100}\right>$ direction.
These results illustrate both how the choice of implantation direction for radiation controls not just the distance at which radiation damage occurs, but also how wide that distance varies between individual projectiles.
Our combination of ML and TD-DFT enables predicting both quantities directly from first principles.

The largest limitation of our simple equation of motion is the inability to model radiation particles that are deflected nor energy transfer to surrounding atoms.
Deflection and energy transfer are important especially when either forces are large (e.g., when the projectile passes close to atoms) or the projectile spends an extend time near a particular atom (e.g., at low velocities).
As such, we expect that the Bragg peaks at low displacements in Fig.~\ref{fig:bragg} are inaccurate because they correspond to trajectories where the projectile passes near atoms. 
Energy transfer to the surrounding atoms only affects the projectile if it is slow enough to still be near the atom after it moves, which occurs for velocities far below where the Bragg peak occurs
\cite{Averback_1998_radiation_damage, Correa:2018}.
Predicting the full force vector on the projectile rather than just the projection of the force along the direction of travel would allow for the deflection of the projectile, and including the forces acting on atoms would unlock accounting for momentum transfer.
Both are directions of future work.

\change{
\subsection*{Extension to Other Materials}

We designed our approach with the goal of extending other materials and projectiles.

The choice of input features will require slight modification for more complex host lattices.
The effect of screening in Coulomb repulsion between the projectile and host lattice must be treated when the repulsion strength of each atom is not equivalent due to symmetry, as in Aluminum.
We propose to fit the effective charges of each symmetric site to forces when projectiles pass near atoms, where the Coulomb forces dominate (see Fig.~\ref{fig:features}).
We would also expand the number features based on AGNI fingerprints, which capture the effect of interactions beyond ionic repulsion, to account for each type of element in a particular material.
Electronic densities are structure and chemistry agnostic, so can be left unchanged.

The above results suggest that only a few trajectories are necessary to train a surrogate for at least simple systems, 
and it may be possible to further reduce the up-front cost for studying a new systems.
The cost for producing the training data used by our surrogate was 31 $\times$ 10\textsuperscript{3} cores-hours on a compute cluster in 2015, and
is likely smaller and even more achievable on on modern hardware.
We hypothesize that the cost can be further reduced by optimal experimental data, such as by halting trajectory calculations as models converge (as visible in Fig.~\ref{fig:time_evolution}).
It may also be possible to further reduce cost by selecting trajectories which optimize diversity of training data, which we explore in the Supplementary Information, or training models on data from multiple materials.
We plan to study these concepts further in ongoing work.
}

\section*{Discussion}

We present a combined TD-DFT and ML route that enables rapid, first-principles prediction of stopping power of materials and the estimation of properties beyond length scales accessible to quantum mechanical simulations.
We demonstrated our method by re-using data from a TD-DFT study on Aluminum to create a model that can predict material properties beyond both the scope of the original study and the present capabilities of TD-DFT (e.g., Bragg Peaks).
This dramatic increase in speed and scope of TD-DFT computations opens a route to producing data that are only rarely available yet critical for technologies including fusion energy, nuclear medicine, and space travel.
Coupled with modern TD-DFT software able to use the latest exascale systems\cite{andrade2021inq}, it is now possible to thoroughly assess electronic stopping in mere hours.



\noindent \textbf{Supplementary Material} accompanies this paper at {\small {\tt http://www.scienceadvances.org/}}.

\section*{Materials and Methods}
%

\subsection*{RT-TDDFT and electronic stopping}

Electronic stopping of a fast projectile, with a velocity on the order of the Bohr velocity, in a target material is a nonadiabatic effect \cite{schleife2015tddftstopping}, which requires a model that can describe electronic excitations caused by fast ionic motion. 
In other words, calculation of electronic stopping requires methods that go beyond the adiabatic Born-Oppenheimer approximation used in static first-principles methods like density functional theory (DFT). 
To address this, we used real-time time-dependent DFT (RT-TDDFT) that is parameter-free and well-demonstrated in the literature\cite{schleife2015tddftstopping,Pruneda_2007_PRL_tddft,Correa:2018,Lee:2020} to capture electronic stopping of fast projectiles. 
The simulation is performed by numerically solving the time-dependent Kohn-Sham (TDKS) equation:
\begin{equation}
    i\frac{\partial}{\partial t} \phi_j(\mathbf{r},t) = \hat{H}[n](t)\,\phi_j(\mathbf{r},t),
    \label{eq:tdks}
\end{equation}
where $\phi_j(\mathbf{r},t)$ are time-dependent Kohn-Sham orbitals and $n$ is the time-dependent electron density.
The Hamiltonian $\hat{H}$,
\begin{equation}
    \label{eq:hamiltonian}
    \hat{H}[n](t)=
    \hat{T} + \hat{V}_{\mathrm{ext}}(t) + \hat{V}_{\mathrm{H}}[n] + \hat{V}_{\mathrm{XC}}[n],
\end{equation}
contains the kinetic energy $\hat{T}$, the external potential $\hat{V}_{\mathrm{ext}}(t)$ due to nuclei and/or external fields, the Hartree electron-electron potential $\hat{V}_{\mathrm{H}}[n]$, and the exchange-correlation potential $\hat{V}_{\mathrm{XC}}[n]$.
Specifically, we used the implementation based on pseudopotential and plane wave basis\cite{schleife:2012}.
For proton-irradiated aluminum studied in this work, we used a simulation cell of 256 atoms with lattice constant of 4.05 \AA\ and the Brillouin zone is sampled only by $\Gamma$ point.
The cutoff energy of 50 Ry were used for the plane wave basis. 
Adiabatic local density approximation was used for exchange-correlation potential and the fourth-order Runge-Kutta scheme, with time step of 0.35 attosecond, was used to numerically time integrate the TDKS equation. 
Schleife et al.\ previously showed that such a simulation setup converges the error in electronic stopping to less than 10\textsuperscript{-2} E$_{H}$/a$_{B}$ for proton-irradiated Aluminum\cite{schleife2015tddftstopping}.

\subsection*{Machine Learning}
We summarize the machine learning techniques below, and provide full details in the Jupyter notebooks and data published along with this study.

\vspace{1ex}
\noindent
\textbf{Data processing and feature computation} are performed by using a combination of the Atomic Simulation Environment (ase) to parse TD-DFT outputs\cite{HjorthLarsen2017ase} and matminer-based featurizers we implemented for this study \cite{ward2018matminer}. Specific details for each feature:

\begin{itemize}
    \item Ion-ion interaction forces are computed by using Ewald summation as implemented in Pymatgen\cite{ong2013pymatgen} assuming a nuclear charges equal to the atomic number.
    \item The local electron density features are determined by interpolating the ground-state electronic density determined from DFT at the projectile's current position as well as the anticipated positions 0.5, 1, 2 time units ahead of and 0.5, 1, 2, 3, 4 time units behind based on the projectile's velocity. 
    \item The directional AGNI fingerprints are \textcolor{red}{projected} along the direction of travel by using Gaussian windows centered at $r=0$ with eight window widths ($\eta$) spaced logarithmically between 0.8 and 16 \AA, inclusive. We used the definition of AGNI fingerprints from Botu et al.\cite{botu2017machine}.

\begin{equation}
    \label{eq:agni}
    V_i^u(\eta) = \sum_{j=i} \frac{r_{ij} \cdot u}{\Vert r_{ij} \Vert} e^{-(\Vert r_{ij} \Vert/\eta)^2}f_d(\Vert r_{ij} \Vert)
\end{equation}

    where $r_{ij}$ is the displacement vector between the projectile ($i$) and a nearby atom ($j$), $u$ is the direction of travel for the projectile, $f_d$ is cutoff function ($f_d (r) = 0.5 \left[ cos(\pi r / R_c) + 1\right]$), and $R_c$ is a cutoff of 16 \AA.
    
\end{itemize}

\vspace{1ex}
\noindent
\textbf{Single-velocity machine learning} models are built by using scikit-learn\change{\cite{scikit-learn}}.
We evaluated Ordinary Least Squares,
LASSO with and without second-degree polynomial features,
Bayesian ridge regression with and without second-degree polynomial features by using
feature selection based on PCA and LASSO.
We used two-fold cross validation to fit hyperparameters where applicable.
\change{Each data point was weighted equally when fitting the the model, regardless of scale.}

\vspace{1ex}
\noindent
\textbf{Multi-velocity machine learning models} are built by using Keras backed by Tensorflow\change{\cite{tensorflow2015-whitepaper}}.
Our chosen model is a multi-layer perceptrons with six hidden dense layers of between 3 and 32 units and
Exponential Linear Unit (ELU) activation functions.
The model was trained by using the Adam optimizer with a learning rate that starts at 5 $\times$ 10\textsuperscript{-4} then decays after the 
validation loss fails to improve for 10 epochs.
\change{The model was trained by using a mean absolute error loss function with all points weighted equally.}

\vspace{1ex}
\noindent
\textbf{Workflows that employ machine learning} rely heavily on numerical routines from SciPy and are parallelized by using Parsl\cite{babuji2019parsl}.
Evaluating the stopping power for a trajectory uses symmetry operations in Pymatgen to determine the 
shortest interval over which to compute the average stopping force,
Pymatgen's nearest-neighbor determination logic to find discontinuities in the stopping force due to 
near passes between projectile and lattice atoms,
and the QUADPACK implementation of Gaussian quadrature to estimate the average stopping force
between those discontinuities.
The stopping distance computations use the fifth-order Runge-Kutta method as implemented in SciPy
to solve Newton's laws of motion given forces computed by using our neural network model \cite{dormand1980rk54}.
Stopping power and distance computations were run across multiple cores of a desktop computer and
multiple nodes of ALCF's Theta supercomputer by using Parsl \cite{babuji2019parsl}.


\bibliography{sciadvbib}
\bibliographystyle{ScienceAdvances}

\noindent \textbf{Acknowledgements:} 
%
The authors thank Kai Nordlund, Flyura Djurabekova, and Andrea Sand for insightful discussions.\\
\noindent \textbf{Funding:}
This work was supported in part by the U.S.\ Department of Energy under contract DE-AC02-06CH11357, and used resources of the Argonne Leadership Computing Facility, a DOE Office of Science User Facility supported under Contract DE-AC02-06CH11357.
A.S.\ and C.-W.L. acknowledge funding by the Office of Naval Research (Grant No.\ N00014-18-1-2605) and the National Science Foundation (Grant Nos.\ OAC-1740219 and OAC-2209857).\\
\noindent \textbf{Author Contributions} All conceptualized the research and contributed to developing methodology; LW and TM implemented software; AS performed validation RT-TDDFT calculations; LW and BB curated data; LW, CL, and AS wrote original draft; all participated in review and editing.\\

\noindent \textbf{Competing Interests} The authors declare that they have no competing financial interests.\\
\noindent \textbf{Data and materials availability:} Additional data and materials are available online.
The datasets include the outputs from RT-TDDFT simulations\cite{2015data,2018data},
and the software and output files from the machine learning study \cite{ward2023sidata}.
The machine learning software are also available on GitHub at
\url{https://github.com/globus-labs/stopping-power-ml/releases/tag/mdfv231027},



\end{document}


\title{Accelerating Electronic Stopping Power Predictions by 10 Million Times with a Combination of Time-Dependent Density Functional
Theory and Machine Learning}

\maketitle

\section{Validation of Single-Velocity Machine Learning Models}

We tested a total of 8 machine learning algorithms to create models of the stopping force acting on a projectile using data from a single velocity.
We only highlight the best-performing model in the paper, but provide all data in Table~\ref{tab:ml_compare}.

\begin{table}[]
    \caption{Comparison of the performance of different variations of linear regression on predicting the stopping force as a function of projectile positions. Mean absolute errors (MAE) are expressed in units of $10^-3$ atomic units of force, $E_H/a_B$. The MAE on small forces is only measured for positions where the true force is below 0.4~$E_H/a_B$.  $\rho^2$ is the squared Spearman's rank correlation coefficient. The best value for each metric is bolded. }
    \centering
    \begin{tabular}{|p{6cm}|c|c|c|}
    \hline Algorithm & MAE, total & MAE, small forces & $\rho^2$ \\ \hline \hline
 Bayesian Linear Regression\newline with LASSO feature selection\newline using Polynomial Features &	25.2 &	\textbf{14.2} &	\textbf{0.987} \\ \hline
LASSO using Polynomial Features	& 30.8 & 14.7 & 0.986 \\ \hline
Bayesian Linear Regression & \textbf{21.9} & 17.4 & 0.980 \\ \hline
LASSO & 39.5 & 19.6 & 0.979 \\ \hline
Bayesian Linear Regression \newline with LASSO feature selection & 40.1 & 22.1 & 0.976 \\ \hline
Bayesian Linear Regression \newline using Polynomial Features & 40.4 & 16.2 & 0.972 \\ \hline
Ordinary Linear Regression & 31.2 & 23.7 & 0.960 \\ \hline
Bayesian Linear Regression \newline with PCA feature selection \newline using Polynomial Features & 47.4 & 21.6  &  0.947 \\ \hline
    \end{tabular}
    \label{tab:ml_compare}
\end{table}

\section{Optimizing Structural Diversity of a Trajectory}

Projectiles traveling through a material along different trajectories can experience greatly different amounts of structural diversity.
A projectile traveling exactly down the center of a high-symmetry channel experiences the same environment every few 
Angstroms of displacement whereas a projectile traveling a random directory never experiences the exact environment twice.
We explored how to quantify the differences in diversity as part of our study and used an optimization strategy to create a trajectory which samples maximal diversity over a specified distance.\footnote{The run parameters used when producing the trajectory described in the main text have been overwritten. We describe the best approximation of the run but cannot guarantee that we have exactly replicated the provenance of that trajectory.}

\subsection{Quantifying Diversity}\label{sec:score}

The goal in quantifying diversity is to measure what fraction of the possible projectile environments are covered by a trajectory.

We define the space of possible projectile environments as all possible positions of a projectile
within the unit cell combined with all directions of travel.
We approximate the space by sampling $2 ^ {14}$ (16384) positions within the primitive cell of Aluminum and sample directions from 
within the unit sphere.

Our next step is to quantify the similarity between projectile environments.
We first assign each point a coordinate using all nine charge density features used for our machine learning models, which capture
the environment both at the projectile's position along with those ahead and behind the projectile.
There is correlation between these coordinates, so we compress them to a 2-dimensional feature space using the IsoMap algorithm.\cite{tenenbaum2000isomap}
We then define the similarity between two points as their distance within the two-dimensional feature space.

We scored the diversity of environments sampled by a trajectory by measuring the fraction of the possible environments the projectile passes near.
We first enumerate points at a timestep of 1 time unit along the trajectory then assign each point to exactly one point from the possible environments.
We then eliminate any these pairs of points which are farther apart than a tolerance threshold.
We chose a threshold distance equal to 5\% of range of coordinates for the sampled set, which corresponds to 1.32 distance units in the space after projecting using ISOMap.
We then count the unique number of points from the sampled space amoung the remaining pairs and divide by the total number generated ($2^{14}$ in our case)
to compute the fraction of possible environments which have been sampled.

\begin{figure}
    \centering
    \includegraphics{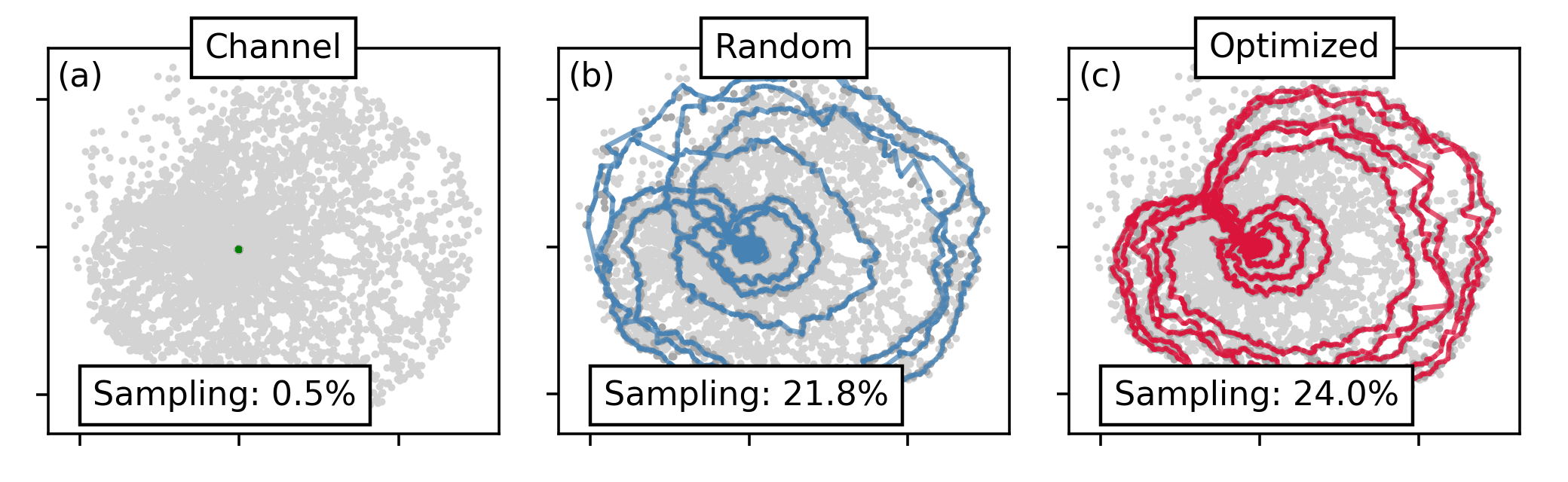}
    \caption{Sampling effectiveness of different projectile trajectories. Each subfigure shows the path of a projectile projected as a colored line and the atomic environments for $2^{14}$ randomly-generated points. The coordinates for the lines and points are determined based on features of the environment (e.g., local charge density) projected into two dimensions using IsoMap.\cite{tenenbaum2000isomap} We show the trajectories for a projectile following (a) the $\left<100\right>$ channel, 
    (b) a randomly-selected path, and (c) a path optimized to sample as many environments as possible. The sampling fraction is determined by measuring the fraction of randomly-generated points close to the trajectory.}
    \label{fig:trajectories}
\end{figure}

The proposed algorithm meets our expectation that the channel trajectory, which repeats the same environment, 
samples fewer environments than the random trajectory.
We measure a 40-fold difference between the two with the channel sampling 0.5\% of all possible environments versus 22.\% for the random trajectory, as shown in Fig.~\ref{fig:trajectories}.

\subsection{Maximizing Diversity}

\begin{figure}
    \centering
    \includegraphics{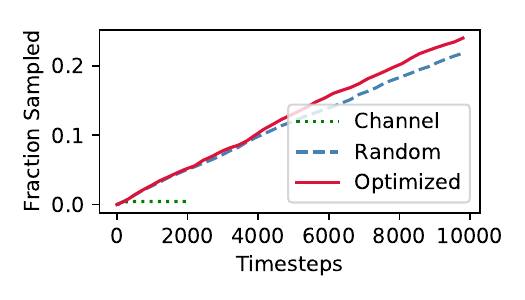}
    \caption{Sampling rate for three different trajectories over time. We compare (a) the $\left<100\right>$ channel, 
    (b) a randomly-selected path, and (c) a path optimized to sample as many environments as possible. The sampling fraction is determined by measuring the fraction of randomly-generated points close to the trajectory.}
    \label{fig:sampling_rate}
\end{figure}

We designed a trajectory with maximal diversity using basis hopping.
The inputs to the optimizer are a starting point of the trajectory in Cartesian coordinates, and the azimuthal and inclination angles for the direction.
The goal of the optimize ris to maximize the environment diversity metric defined in \S\ref{sec:score}.
We used the starting position and direction for the random trajectory for an initial guess and kept the number of points and velocity in the trajectory fixed (9800 steps, $v=1$ a.u.).
We ran the optimizer for 100 steps and yielded a trajectory which sampled 24\% of the total space.

Like the random trajectory, the \textit{optimized} trajectory samples new environments at a consistent rate over time but the rate is larger for the optimized trajectory.
The increased sampling performance comes at a cost of increased bias, which could be problematic in training a machine learning model, but does provide a unique test case for the model.
Determining how best to use these optimized trajectories in training would be an interesting line for future research.

\bibliography{sciadvbib}
\bibliographystyle{ScienceAdvances}